# Information sets from defining sets in abelian codes

José Joaquín Bernal and Juan Jacobo Simón

*Abstract*—We describe a technique to construct a set of check positions (and hence an information set) for every abelian code solely in terms of its defining set. This generalizes that given by Imai in [7] in the case of binary TDC codes.

## I. Introduction

In the study of any family of error-correcting codes, information sets are essential ingredients for encoding and decoding purposes and hence it is important to describe effective algorithms for finding them. This is the interest of this paper for the family of abelian codes [1], [3].

In [7], Imai gave a method to obtain information sets for binary two dimensional cyclic (TDC) codes. Later, Sakata [12] gave an alternative method for the same purpose. Imai's algorithm relies in the structure of the roots of the code while the algorithm of Sakata is somehow based on the division algorithm for polynomials. Up to our knowledge, these are the only techniques available for TDC codes. Following the ideas in the papers mentioned above, Chabanne described a method to calculate syndromes via Groebner basis and then he generalized the usual permutation decoding procedure (see [4]).

In this paper, we generalize Imai's method to arbitrary abelian codes. As it was done by Imai, our method is based on computation of cardinalities of cyclotomic cosets on different extensions of the ground field. Such cosets, as well as their cardinalities, are completely determined by the structure of the defining set of the code (see below for definitions). It turns up that these cosets determine directly the shape of the information set. This relationship allows us to construct codes with prescribed information sets in order to get particular properties; or, in the case of a fixed code, allows us to determine, *a priori*, the shape of its information set. This takes an advantage with respect to the use of the Groebner basis to obtain them. As an application, we shall design abelian codes with good information sets in order to implement the permutation decoding algorithm (see [9]). Indeed, we obtain codes with better rates that those obtained with other information sets.

This paper is devoted to study a construction of sets of check positions for the family of abelian codes. In Section III we describe a method to construct information sets for abelian codes solely in terms of their defining set. In Section IV we include some examples which complete the exposition of our construction. In Section V we state our main result; to witt, that our set is in fact a set of check positions. In Section VI we show how one may design abelian codes with suitable

Departamento de Matemáticas, Universidad de Murcia, 30100 Murcia. Spain. email: josejoaquin.bernal@um.es, jsimon@um.es
Partially supported by D.G.I. of Spain and Fundación Séneca of Murcia.

information sets in order to make a permutation decoding attempt.

An extended abstract of this paper, entitled "Information sets for abelian codes", appears in the proceedings of 2010 IEEE Information Theory Workshop, Dublin, august 30-september 3, 2010.

## II. Preliminaries

All throughout $\mathbb{F}$ denotes the field with $q$ elements where $q$ is a power of a prime $p$. Let $\mathcal{C}$ be a linear code of dimension $k$, and length $l$ over the field $\mathbb{F}$. An information set for $\mathcal{C}$ is a set of positions $\{i_1, \dots, i_k\}$ such that restricting the codewords to these positions we get the whole space $\mathbb{F}^k$; the other $l - k$ positions are called check positions (see [10], [11]).

A (left) group code of length $l$ is a linear code which is the image of a (left) ideal of a group algebra via an $\mathbb{F}$-isomorphism $\mathbb{F}G \to \mathbb{F}^l$ which maps $G$ to the standard basis of $\mathbb{F}^l$. If $\mathcal{G}$ is a family of groups, we say that a linear code is a (left) $\mathcal{G}$ group code if it is a (left) group code for some group $G$ belonging to $\mathcal{G}$ (see [2]). Many classical linear codes have been shown to be group codes: cyclic codes, Reed-Muller codes, etc. We deal with abelian codes, that is, $\mathcal{G}$ group codes where $\mathcal{G}$ is the class of finite abelian groups.

We recall some basic facts about abelian codes (see [1], [3] for details). Every abelian code may be identified with an ideal of a group algebra $\mathbb{F}G$, where $G$ is an abelian group. It is well-known that a decomposition $G \simeq C_{r_1} \times \cdots \times C_{r_n}$, with $C_{r_i}$ the cyclic group of order $r_i$, induces a canonical isomorphism of algebras $\mathbb{F}G \simeq \mathbb{F}[X_1, \dots, X_n] / \langle X_1^{r_1} - 1, \dots, X_n^{r_n} - 1 \rangle$. We denote this quotient algebra by $\mathbb{A}(r_1, \dots, r_n)$. Then every abelian code may be viewed as an ideal of $\mathbb{A}(r_1, \dots, r_n)$, and so we identify the codewords with polynomials $P(X_1, \dots, X_n)$ such that every monomial satisfy that the degree of the variable $X_i$ is in $\mathbb{Z}_{r_i}$, the cyclic group of non negative integers less or equal than $r_i$. We write the elements $P(X_1, \dots, X_n) \in \mathbb{A}(r_1, \dots, r_n)$ as $P(X_1, \dots, X_n) = \sum a_{\mathbf{j}} X^{\mathbf{j}}$, where $\mathbf{j} = (j_1, \dots, j_n) \in \mathbb{Z}_{r_1} \times \cdots \times \mathbb{Z}_{r_n}$ and $X^{\mathbf{j}} = X_1^{j_1} \cdots X_n^{j_n}$.

We always assume that $\gcd(r_i, q) = 1$ for every $i = 1, \dots, n$. Our construction make use of the structure of roots of the ideals in $\mathbb{A}(r_1, \dots, r_n)$; so let us recall some basic facts about it. We fix a primitive $r_i$-th root of unity $\alpha_i$ in some extension of $\mathbb{F}$, for each $i = 1, \dots, n$. It is well known that every abelian code $\mathcal{C} \leq \mathbb{A}(r_1, \dots, r_n)$ is totally determined by its *root* set, defined as

$$\mathcal{Z}(\mathcal{C}) = \{(\alpha_1^{a_1}, \dots, \alpha_n^{a_n}) \mid P(\alpha_1^{a_1}, \dots, \alpha_n^{a_n}) = 0$$
$$\text{for all } P(X_1, \dots, X_n) \in \mathcal{C}\}.$$

As it is usual, we consider the *defining* set of $\mathcal{C}$, to witt,

$$D(\mathcal{C}) = \{(a_1, \dots, a_n) \in \mathbb{Z}_{r_1} \times \cdots \times \mathbb{Z}_{r_n} \mid$$
$$(\alpha_1^{a_1}, \dots, \alpha_n^{a_n}) \in \mathcal{Z}(\mathcal{C})\}.$$

Given an abelian code $\mathcal{C} \leq \mathbb{A}(r_1, \dots, r_n)$ with defining set $D(\mathcal{C})$ with respect to $\{\alpha_i\}_{i=1}^n$, if one chooses different primitive roots of unity then the set $D(\mathcal{C})$ detemines a new code, say $\mathcal{C}'$, which is permutation equivalent to $\mathcal{C}$. So, for the sake of shortness, we refer to abelian codes without any mention to the primitive roots that we are using as reference.

Recall that, for $\gamma \in \mathbb{N}$, the $q^\gamma$-cyclotomic coset of an integer $a$ modulo $r$ is the set

$$C_{(q^\gamma, r)}(a) = \left\{ a \cdot q^{\gamma \cdot i} \mid i \in \mathbb{N} \right\} \subseteq \mathbb{Z}_r.$$

We extend the concept of $q$-cyclotomic coset modulo an integer to "$q$-cyclotomic coset modulo a tuple of integers". Given an element $(a_1, \ldots, a_n) \in \mathbb{Z}_{r_1} \times \cdots \times \mathbb{Z}_{r_n}$, we define its $q$-orbit modulo $(r_1, \ldots, r_n)$ as

$$
\begin{aligned}
Q(a_1, \ldots, a_n) &= \left\{ (a_1 \cdot q^i, \ldots, a_n \cdot q^i) \mid i \in \mathbb{N} \right\} \quad (1) \\
&\subseteq \mathbb{Z}_{r_1} \times \cdots \times \mathbb{Z}_{r_n}.
\end{aligned}
$$

It is easy to check that

$$|Q(a_1, \ldots, a_n)| = \operatorname{lcm}\left( \left| C_{(q, r_1)}(a_1) \right|, \ldots, \left| C_{(q, r_n)}(a_n) \right| \right).$$

For the sake of shortness we only write $q$-orbit, and the tuple of integers $(r_1, \ldots, r_n)$ will be clear by context.

It is also easy to see that $D(\mathcal{C})$ is closed under multiplication by $q$ in $\mathbb{Z}_{r_1} \times \cdots \times \mathbb{Z}_{r_n}$, so, one has that $D(\mathcal{C})$ is necessarily a disjoint union of $q$-orbits.

Conversely, every union of $q$-orbits modulo $(r_1, \ldots, r_n)$ defines an abelian code in $\mathbb{A}(r_1, \ldots, r_n)$.

## III. Construction of information sets

In this section we shall describe a method to construct sets of check positions for abelian codes in terms of its defining set. We assume all throughout that $n > 1$. We will comment the case $n = 1$ (cyclic codes) in Remark 2 (c).

Let $\mathcal{C} \leq \mathbb{A}(r_1, \ldots, r_n)$ be an abelian code with defining set $D(\mathcal{C})$. For any element $e \in D(\mathcal{C})$ and for each $1 \leq i \leq n$ let $\pi_i(e)$ denote the projection onto the first $i$-coordinates. We set $D_i(\mathcal{C}) = \pi_i(D(\mathcal{C}))$; moreover, for each $1 \leq t \leq i$ and $e \in D_i(\mathcal{C})$ we also denote by $\pi_t(e)$ its projection onto the first $t$-coordinates.

We fix a well-ordering in the indeterminates setting $X_1 < \cdots < X_n$, and we define recursively the following parameters. For each $e = (e_1, \ldots, e_j) \in D_j(\mathcal{C})$, with $1 \leq j \leq n$, we set $m(\pi_1(e)) = \left| C_{(q, r_1)}(e_1) \right|$, and having $m(\pi_t(e))$ defined for $1 \leq t \leq j - 1$, we set

$$\gamma_{t+1}(e) = \prod_{l=1}^{t} m(\pi_l(e))$$

and

$$m(\pi_{t+1}(e)) = \left| C_{(q^{\gamma_{t+1}(e)}, r_{t+1})}(e_{t+1}) \right|.$$

Finally, for $e \in D(\mathcal{C})$, we set $\gamma_{n+1}(e) = \prod_{l=1}^{n} m(\pi_l(e))$. It is easy to see that for $1 \leq t \leq n$, we have that $\gamma_{t+1}(e) = |Q(\pi_t(e))|$ and for $1 \leq t < n$, one has $m(\pi_t(e)) = \gamma_{t+1}(e)/\gamma_t(e)$.

**Remark 1.** *Given a $q$-orbit, any of its representatives yields the same parameters $m$'s and $\gamma$'s.*

Let us prove this assertion. We argument by induction. The case $n = 1$ is the obvious statement: for $a, a' \in \mathbb{Z}_r$, if $C_{(q, r)}(a) = C_{(q, r)}(a')$ then $\left| C_{(q^\gamma, r)}(a) \right| = \left| C_{(q^\gamma, r)}(a') \right|$ for any $\gamma \in \mathbb{N}$. Suppose that $m(\pi_i(e)) = m(\pi_i(e'))$ for all

$1 \leq i \leq t < n$, then $\gamma_{t+1}(e) = \gamma_{t+1}(e')$ and so, by the case $n = 1$, $\left| C_{(q^{\gamma_{t+1}(e)}, r_{t+1})}(e_{t+1}) \right| = \left| C_{(q^{\gamma_{t+1}(e')}, r_{t+1})}(e'_{t+1}) \right|$ which implies $m(\pi_{t+1}(e)) = m(\pi_{t+1}(e'))$, as desired. ∎

Using the previous fixed ordering we begin our construction by choosing a set of representatives of the $q$-orbits, which we will denote by $\overline{D}(\mathcal{C})$, verifying the following restriction: if $e = (e_1, \ldots, e_n)$ and $e' = (e'_1, \ldots, e'_n)$ are elements of $\overline{D}(\mathcal{C})$ such that $\gamma_t(e) = \gamma_t(e')$ and $e_t, e'_t$ belong to the same $q^{\gamma_t(e)}$-cyclotomic coset modulo $r_t$, for some $t \in \{1, \ldots, n\}$, then it must happen $e_t = e'_t$. We call $\overline{D}(\mathcal{C})$ a set of restricted representatives. As we will see later, every defining set $D(\mathcal{C})$ has at least one subset of restricted representatives. (The reader may see the role of this restriction in Example 6.)

In what follows, the previous ordering $X_1 < \cdots < X_n$ will be fixed as default order. We are denoting by $\Gamma(\mathcal{C})$ the construction below [see (2)] in which we use the default ordering. When we use alternative orderings we will make adequate notational changes (see Example 5).

For every $i \in \{1, \ldots, n-1\}$, let us denote by $\overline{D}_i(\mathcal{C})$ the image of the projection of $\overline{D}(\mathcal{C})$ onto the first $i$-coordinates, and given $e \in \overline{D}_i(\mathcal{C})$, let

$$R(e) = \{ a \in \mathbb{Z}_{r_{i+1}} \mid (e, a) \in \overline{D}_{i+1}(\mathcal{C}) \},$$

where $(e, a)$ has the obvious meaning; that is, if $e = (e_1, \ldots, e_i)$ then $(e, a) = (e_1, \ldots, e_i, a)$.

For each $e \in \overline{D}_{n-1}(\mathcal{C})$, we define

$$M(e) = \sum_{a \in R(e)} m(e, a)$$

and consider the set $\{M(e)\}_{e \in \overline{D}_{n-1}(\mathcal{C})}$. Then we denote the different values of the $M(e)$'s as follows,

$$
\begin{aligned}
f[1] &= \max_{e \in \overline{D}_{n-1}(\mathcal{C})} \{M(e)\} \quad \text{and} \\
f[i] &= \max_{e \in \overline{D}_{n-1}(\mathcal{C})} \{M(e) \mid M(e) < f[i-1]\}.
\end{aligned}
$$

So, we obtain the sequence

$$f[1] > \cdots > f[s_n] > 0 = f[s_n + 1],$$

where $f[s_n + 1] = 0$ by convention.

This is the initial sequence for any value of $n$. Now, suppose that $n \geq 3$. Then we continue as follows:

Given $1 \leq u_n \leq s_n$, we define for every $e \in \overline{D}_{n-2}(\mathcal{C})$

$$\Omega_{u_n}(e) = \{ a \in R(e) \mid M(e, a) \geq f[u_n] \}$$

and

$$\mu_{u_n}(e) = \sum_{a \in \Omega_{u_n}(e)} m(e, a).$$

Note that the set $\Omega_{u_n}(e)$ may eventually be the empty set. In this case, the corresponding value $\mu_{u_n}(e)$ will be zero.

We define

$$
\begin{aligned}
f[u_n, 1] &= \max_{e \in \overline{D}_{n-2}(\mathcal{C})} \{\mu_{u_n}(e)\} \quad \text{and} \\
f[u_n, i] &= \max_{e \in \overline{D}_{n-2}(\mathcal{C})} \{\mu_{u_n}(e) \mid 0 < \mu_{u_n}(e) < f[u_n, i-1]\}.
\end{aligned}
$$

For each $1 \leq u_n \leq s_n$ we order the previous parameters getting the sequence

$$f[u_n, 1] > \cdots > f[u_n, s(u_n)] > 0 = f[u_n, s(u_n) + 1],$$

where, again, $f[u_n, s(u_n) + 1] = 0$ by convention.

Now, suppose that $n \geq j \geq 4$ and we have constructed for each list $(u_n, \ldots, u_j)$, where $1 \leq u_n \leq s_n$ and $1 \leq u_l \leq s(u_n, \ldots, u_{l+1})$, for $j \leq l < n$, the sequence

$$f[u_n, \ldots, u_j, 1] > \cdots > f[u_n, \ldots, u_j, s(u_n, \ldots, u_j)] > 0 =$$
$$= f[u_n, \ldots, u_j, s(u_n, \ldots, u_j) + 1].$$

Then we define for each $e \in \overline{D}_{j-3}(\mathcal{C})$ and $1 \leq u_{j-1} \leq s(u_n, \ldots, u_j)$

$$\Omega_{u_n, \ldots, u_{j-1}}(e) = \{ a \in R(e) \mid \mu_{u_n, \ldots, u_j}(e, a) \geq f[u_n, \ldots, u_j, u_{j-1}] \}$$

and

$$\mu_{u_n, \ldots, u_{j-1}}(e) = \sum_{a \in \Omega_{u_n, \ldots, u_{j-1}}(e)} m(e, a),$$

and by ordering these parameters we obtain the sequence

$$f[u_n, \ldots, u_{j-1}, 1] > \cdots > f[u_n, \ldots, u_{j-1}, s(u_n, \ldots, u_{j-1})]$$
$$> 0 = f[u_n, \ldots, u_{j-1}, s(u_n, \ldots, u_{j-1}) + 1].$$

We continue in this way until we have defined, for each $(u_n, \ldots, u_3)$, the sequence

$$f[u_n, \ldots, u_3, 1] > \cdots > f[u_n, \ldots, u_3, s(u_n, \ldots, u_3)] > 0 =$$
$$= f[u_n, \ldots, u_3, s(u_n, \ldots, u_3) + 1].$$

Finally, for any value of $n$, we define the last list of parameters as follows: for every list $(u_n, \ldots, u_2)$, with $1 \leq u_n \leq s_n$ and $1 \leq u_l \leq s(u_n, \ldots, u_{l+1})$, with $2 \leq l < n$, if $n = 2$ we set

$$g[u_2] = \sum_{\substack{e \in \overline{D}_1(\mathcal{C}) \\ M(e) \geq f[u_2]}} m(e)$$

and, in case $2 < n$, we set

$$g[u_n, \ldots, u_2] = \sum_{\substack{e \in \overline{D}_1(\mathcal{C}) \\ \mu_{u_n, \ldots, u_3}(e) \geq f[u_n, \ldots, u_2]}} m(e).$$

Using all the sequences obtained previously, we may define the set

$$\Gamma(\mathcal{C}) = \left\{ (i_1, \ldots, i_n) \in \prod_{i=1}^n \mathbb{Z}_{r_i} \mid \exists (u_n, \ldots, u_2) \right.$$
$$\text{where} \tag{2}$$
$$1 \leq u_n \leq s_n, \text{ and}$$
$$1 \leq u_l \leq s(u_n, \ldots, u_{l+1}), \text{ for } l = 2, \ldots, n-1$$
$$\text{such that}$$
$$f[u_n + 1] \leq i_n < f[u_n],$$
$$\cdots$$
$$f[u_n, \ldots, u_2 + 1] \leq i_2 < f[u_n, \ldots, u_2],$$
$$\left. 0 \leq i_1 < g[u_n, \ldots, u_2] \right\}.$$

The main result of this paper stays that $\Gamma(\mathcal{C})$ is a set of check positions for $\mathcal{C}$ (see Theorem 9) or equivalently, the complement of $\Gamma(\mathcal{C})$ in $\mathbb{Z}_{r_1} \times \cdots \times \mathbb{Z}_{r_n}$ is an information set of $\mathcal{C}$.

**Remarks 2.** a) In the case $n = 2$ the reader may check that the set

$$\Gamma(\mathcal{C}) = \{ (i_1, i_2) \in \mathbb{Z}_{r_1} \times \mathbb{Z}_{r_2} \mid \exists u \in \{1, \ldots, s_n\}$$
$$\text{with} \tag{3}$$
$$f[u + 1] \leq i_2 < f[u] \text{ and } 0 \leq i_1 < g[u] \}$$

is that constructed by Imai for TDC codes in [7], viewed under our notation.

b) The construction of $\Gamma(\mathcal{C})$ depends uniquely on the values $m(-)$ computed over the restricted sets of representatives, and, as we have seen in Remark 1, this values are independent of the representatives, provided that their election respects the restriction imposed. Therefore, any election of restricted representatives $\overline{D}(\mathcal{C})$ yields the same set of check positions as far as the order of the indeterminates is fixed. However, the values $m(-)$ depend on the ordering of the indeterminates, so that different orderings may produce different sets of check positions, as it is shown in Example 5.

c) *Cyclic codes.* Let $\mathcal{C}$ be a cyclic code of length $l$ over $\mathbb{F}$ with root set $\mathcal{Z}(\mathcal{C})$ and defining set $D(\mathcal{C}) = \{ e \in \mathbb{Z}_l \mid \alpha^e \in \mathcal{Z}(\mathcal{C}) \}$, where $\alpha$ is a primitive $l$-th root of unity. Assume that $\gcd(l, q) = 1$. Let $\overline{D}(\mathcal{C})$ be a complete set of representatives of the $q$-cyclotomic cosets modulo $l$ of the elements of $D(\mathcal{C})$. The specialization for $n = 1$ of the construction given above yields the set of check positions

$$\Gamma(\mathcal{C}) = \{ i_1 \in \mathbb{Z}_l \mid 0 \leq i_1 < \sum_{e \in \overline{D}(\mathcal{C})} m(e) \};$$

that is, the last $l - |D(\mathcal{C})|$ positions define an information set, as it is well known (see [10]).

Now suppose that $l = r_1 \cdot r_2$ with $\gcd(r_1, r_2) = 1$. Then there exists a group isomorphism $\varphi : C_l \to C_{r_1} \times C_{r_2}$ which can be extended by linearity to an isomorphism $\overline{\varphi}$ of group algebras from $\mathbb{F}C_l$ to $\mathbb{F}(C_{r_1} \times C_{r_2})$. Moreover, fixed $\alpha_1$, $\alpha_2$ primitive $r_1$-th, $r_2$-th roots of unity, $\varphi$ maps $D(\mathcal{C})$ onto the set $D(\overline{\varphi}(\mathcal{C}))$. So, we can compute an information set of $\overline{\varphi}(\mathcal{C})$ and then take the inverse image of those positions by $\varphi$, obtaining a new information set for $\mathcal{C}$.

It is not hard to see that different isomorphisms may produce different sets of check positions for $\mathcal{C}$. Indeed, if $\varphi, \varphi' : C_l \to C_{r_1} \times C_{r_2}$ are different group isomorphisms then being $D(\overline{\varphi}(\mathcal{C}))$ and $D(\overline{\varphi'}(\mathcal{C}))$ different or not, the sets of check positions coincide; that is $\Gamma(\overline{\varphi}(\mathcal{C})) = \Gamma(\overline{\varphi'}(\mathcal{C}))$. This is because group isomorphisms preserve the structure of cyclotomic cosets (and so the values $m(-)$). Thus, as $\varphi \neq \varphi'$, their inverses may produce different information sets for $\mathcal{C}$.

Let us give briefly an example. Let $\mathcal{C}$ be the binary cyclic code of length 15 with root set $\mathcal{Z}(\mathcal{C}) = \{ \alpha^e \mid e = 0, 1, 2, 3, 4, 6, 8, 9, 12 \}$ where $\alpha$ is a primitive 15-th root of unity. Using the Chinese Remainder Theorem we define the group isomorphism $\varphi : C_{15} \to C_3 \times C_5$ given by $\varphi(t) = (t_1, t_2)$, where $t_1 \equiv t \mod 3$ and $t_2 \equiv t \mod 5$ are suitable representatives.

Then, considering $\alpha_1$ and $\alpha_2$ primitive third and fifth roots of unity, respectively, such that $\alpha = \alpha_1 \alpha_2$, we obtain

$$\mathcal{Z}(\overline{\varphi}(\mathcal{C})) = \{(1,1),(1,\alpha_2),(1,\alpha_2^2),(1,\alpha_2^3),(1,\alpha_2^4),$$
$$(\alpha_1,\alpha_2),(\alpha_1^2,\alpha_2^2),(\alpha_1,\alpha_2^4),(\alpha_1^2,\alpha_2^3)\}$$

and

$$D(\overline{\varphi}(\mathcal{C})) = \{(0,0),(0,1),(0,2),(0,3),(0,4),(1,1),$$
$$(2,2),(1,4),(2,3)\}.$$

Using our construction we get the set of check positions for $\varphi(\mathcal{C})$,

$$\Gamma(\overline{\varphi}(\mathcal{C})) = \{(0,0),(0,1),(0,2),(0,3),(0,4),(1,0),$$
$$(2,0),(1,1),(2,1)\}.$$

Going back we obtain the set of check positions for $\mathcal{C}$,

$$\varphi^{-1}(\Gamma(\overline{\varphi}(\mathcal{C}))) = \{0,1,3,5,6,9,10,11,12\};$$

a non obvious one. The reader may check that the isomorphism $\varphi'(t) = (3-t_1, t_2)$, with $t_1, t_2$ as above, determines another different set of check positions for $\mathcal{C}$.

It is easy to see that this can be generalized to the case $l = r_1 \cdots r_n$ with pairwise coprime $r_i$'s.

## IV. EXAMPLES

Before proving the fact that $\Gamma(\mathcal{C})$ is a set of check positions of $\mathcal{C}$, we present some examples to illustrate our construction. Let $\mathcal{C}$ be an abelian code with defining set $D(\mathcal{C})$. To construct a set of check positions we first need a set of restricted representatives of the $q$-orbits of $D(\mathcal{C})$; that is, $\overline{D}(\mathcal{C})$. We propose the following inductive algorithm using the default ordering $X_1 < \cdots < X_n$.

- Take a complete set of representatives of the $q$-cyclotomic cosets modulo $r_1$ of the elements of $D_1(\mathcal{C})$. This set will be $\overline{D}_1(\mathcal{C})$, and for the sake of simplicity we use the same symbol.

- Suppose that we have defined $\overline{D}_{i-1}(\mathcal{C})$. Given $e \in \overline{D}_{i-1}(\mathcal{C})$ we define $\overline{D}_i(e)$ as the set of the elements $(e, a_i)$ where $a_i$ runs through a complete set of representatives of the $q^{\gamma_i(e)}$-cyclotomic cosets modulo $r_i$ of the elements of $D_i(e) = \{a \in \mathbb{Z}_{r_i} \mid (e, a) \in D_i(\mathcal{C})\}$. One should have in mind that if $e, e' \in \overline{D}_{i-1}(\mathcal{C})$ and $\gamma_i(e) = \gamma_i(e')$ then we must take the same representatives for each $q^{\gamma_i(e)}$-cyclotomic coset in $D_i(e) \cap D_i(e')$. Set $\overline{D}_i(\mathcal{C}) = \bigcup_{e \in \overline{D}_{i-1}(\mathcal{C})} \overline{D}_i(e)$.

- Finally, we define $\overline{D}(\mathcal{C}) = \overline{D}_n(\mathcal{C})$.

Clearly, any set of restricted representatives, $\overline{D}(C)$, comes from elections in the given way.

**Example 3.** Let $q = 2$, $n = 2$, $r_1 = 3$, $r_2 = 7$, and consider the abelian code $\mathcal{C}_1$ with defining set

$$D(\mathcal{C}_1) = \{(1,1),(2,2),(1,4),(2,1),(1,2),(2,4),$$
$$(0,3),(0,5),(0,6),(1,3),(2,6),(1,5),$$
$$(2,3),(1,6),(2,5)\}.$$

Take $\overline{D}_1(\mathcal{C}) = \{0,1\}$ a complete set of representatives of the 2-cyclotomic cosets modulo 3 of the elements of $D_1(\mathcal{C})$. Then we may take $\overline{D}_2(0) = \{(0,3)\}$ and $\overline{D}_2(1) = \{(1,1),(1,3)\}$. By definition, $\overline{D}(\mathcal{C}_1) = \{(0,3),(1,1),(1,3)\}$. A direct computation yields $m(0) = 1$, $m(0,3) = m(1,1) = m(1,3) = 3$, and $m(1) = 2$. Hence $M(0) = 3$ and $M(1) = 6$. Then, we obtain the (unique) sequence

$$f[1] = 6 > f[2] = 3 > 0 = f[3]$$

and

$$g[1] = 2 < g[2] = 3.$$

Finally, according to (3) the set of check positions for $\mathcal{C}_1$ is

$$\Gamma(C_1) = \{(0,0),(1,0),(2,0),(0,1),(0,2),(0,3),$$
$$(0,4),(0,5),(1,1),(2,1),(1,2),(2,2),$$
$$(1,3),(1,4),(1,5)\}$$

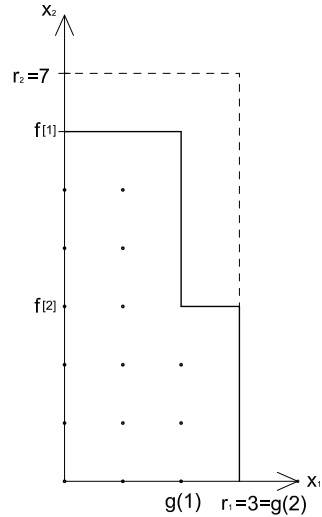

Fig. 1. $\Gamma(\mathcal{C}_1)$

**Example 4.** Set $q = 2, n = 3, r_1 = r_2 = r_3 = 3$. Let $\mathcal{C}_2$ be the abelian code with defining set

$$D(\mathcal{C}_2) = \{(0,0,0),(1,1,0),(2,2,0),(0,1,1),(0,2,2),$$
$$(2,2,1),(1,1,2)\}.$$

Take $\overline{D}_1(\mathcal{C}) = \{0,1\}$ as a complete set of representatives of the 2-cyclotomic cosets modulo 3 of the elements of $D_1(\mathcal{C})$. Choose $\overline{D}_2(0) = \{(0,0),(0,1)\}$, $\overline{D}_2(1) = \{(1,1)\}$. By definition, $\overline{D}_2(\mathcal{C}) = \{(0,0),(0,1),(1,1)\}$. Now, we set $\overline{D}_3(0,0) = \{(0,0,0)\}$, $\overline{D}_3(0,1) = \{(0,1,1)\}$, and $\overline{D}_3(1,1) = \{(1,1,0),(1,1,2)\}$. Then

$$\overline{D}(\mathcal{C}_2) = \{(0,0,0),(0,1,1),(1,1,0),(1,1,2)\}.$$

A direct computation shows that $m(0) = m(0,0) = m(1,1) = m(0,0,0) = m(0,1,1) = m(1,1,0) = m(1,1,2) = 1$, and $m(1) = m(0,1) = 2$. Therefore $M(0,0) = M(0,1) = 1, M(1,1) = 2$ which, in turn, give us the first sequence

$$f[1] = 2 > f[2] = 1 > 0 = f[3].$$

For $u_3 = 1$ we have the values $\mu_1(0) = 0, \mu_1(1) = 1$. This yields a sequence

$$f[1,1] = 1 > 0 = f[1,2].$$

For $u_3 = 2$ the values are $\mu_2(0) = 3, \mu_2(1) = 1$. Then,

$$f[2,1] = 3 > f[2,2] = 1 > 0 = f[2,3].$$

and finally

$$g[1,1] = 2 \text{ and } g[2,1] = 1 < g[2,2] = 3.$$

Following (2), the set of check positions for $\mathcal{C}$ is

$$\Gamma(\mathcal{C}_2) = \{(0,0,0),(1,0,0),(2,0,0),(0,1,0),\\(0,2,0),(0,0,1),(1,0,1)\}.$$

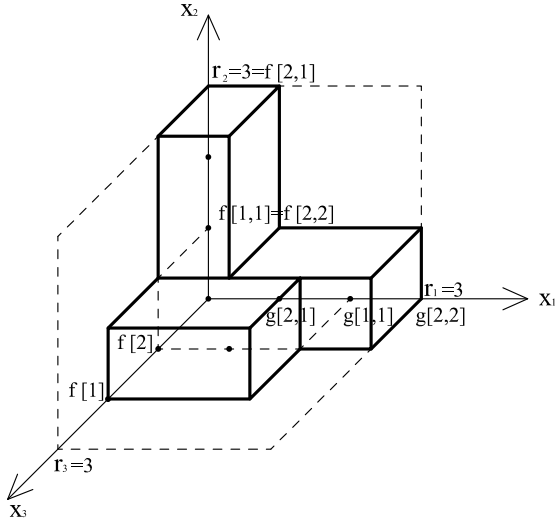

Fig. 2. $\Gamma(\mathcal{C}_2)$

Next example shows that different linear orderings in the indeterminates may yield different sets of check positions for an abelian code.

**Example 5.** Let $q = 2$, $n = 2$, $r_1 = 3$ and $r_2 = 5$. Let $\mathcal{C}_3$ be the abelian goup code with defining set

$$D(\mathcal{C}_3) = \{(0,0),(1,0),(2,0),(1,2),(2,4),(1,3),(2,1)\}.$$

First we choose the (default) order $X_1 < X_2$. We take

$$\overline{D}(\mathcal{C}_3) = \{(0,0),(1,0),(1,2)\}.$$

Then $m(0) = 1$, $m(1) = 2$, $m(0,0) = 1$, $m(1,0) = 1$, $m(1,2) = 2$. Using this, we have $M(0) = 1$, $M(1) = 3$. So we obtain the sequence

$$f[1] = 3 > f[2] = 1 > f[3] = 0$$

and

$$g[1] = 2 < g[2] = 3.$$

Therefore, the set of check positions for $\mathcal{C}_3$ with respect to this order is

$$\Gamma(\mathcal{C}_3; X_1 < X_2) = \{(0,0),(1,0),(2,0),(0,1),(0,2),\\(1,1),(1,2)\}.$$

Now we choose the other order, $X_2 < X_1$. We take

$$\overline{D}(\mathcal{C}_3) = \{(0,0),(1,0),(2,1)\}.$$

Then $m(0) = 1$, $m(1) = 4$, $m(0,0) = 1$, $m(1,0) = 2$, $m(2,1) = 1$. Using this, we have $M(0) = 3$, $M(1) = 1$. So we obtain the sequence

$$f[1] = 3 > f[2] = 1 > f[3] = 0$$

and

$$g[1] = 1 < g[2] = 5.$$

Therefore, the set of check positions for $\mathcal{C}_3$ with respect to the order $X_2 < X_1$ is

$$\Gamma(\mathcal{C}_3; X_2 < X_1) = \{(0,0),(1,0),(2,0),(0,1),(0,2),\\(0,3),(0,4)\}.$$

Our last example shows that the restriction on the election of representatives is not superfluous, and allows us to avoid redundance on the sums.

**Example 6.** Set $q = 2, n = 3, r_1 = r_2 = r_3 = 3$. Let $\mathcal{C}_4$ be the abelian code with defining set

$$D(\mathcal{C}_4) = \{(0,0,0),(0,1,1),(0,2,2),(0,2,1),(0,1,2)\}.$$

One may choose $\overline{D}(\mathcal{C}_4) = \{(0,0,0),(0,1,1),(0,1,2)\}$ or, alternatively, $\overline{D}(\mathcal{C}_4) = \{(0,0,0),(0,2,1),(0,2,2)\}$; but we must not select as $\overline{D}(\mathcal{C}_4)$ the set $\{(0,0,0),(0,1,1),(0,2,1)\}$. If we do this last election, we will obtain $\mu_1(0) = m(0,0) + m(0,1) + m(0,2) = 5$, which makes no sense, because $r_i = 3$, for $i = 1, 2, 3$. Note that the election of $(0,1)$ and $(0,2)$ as an $\overline{D}_2(\mathcal{C})$ does not respect the rules of restricted representatives.

## V. Main theorem

In this section we will prove that $\Gamma(\mathcal{C})$ is a set of check positions. Firstly, we need to introduce the notion of *check tensor* associated with an abelian code $\mathcal{C}$. This definition is an extension to $n$ indeterminates of that given by Imai in [7].

Let $\mathcal{C} \leq \mathbb{A}(r_1, \ldots, r_n)$ be an abelian code with defining set $D(\mathcal{C})$. Suppose we have ordered $X_1 < \ldots < X_n$ obtaining $\overline{D}(\mathcal{C})$ and $\Gamma(\mathcal{C})$ as it was done in Section III.

Let $\mathbb{F}_{q'}$ denote the field of $q'$ elements with $q'$ a power of $q$ (recall from the Preliminaries that $\mathbb{F} = \mathbb{F}_q$). For every $\mathbf{j} = (j_1, \ldots, j_n) \in \mathbb{Z}_{r_1} \times \cdots \times \mathbb{Z}_{r_n}$, and every $e = (e_1, \ldots, e_n) \in \overline{D}(\mathcal{C})$ we define $h_e^{\mathbf{j}}$ as the expression of $\alpha_1^{e_1 \cdot j_1} \cdot \ldots \cdot \alpha_n^{e_n \cdot j_n}$ in a fixed basis of the $\mathbb{F}_q$-vector space $\mathbb{F}_{q'}$, where $q' = q^{\gamma_{n+1}(e)}$ and $\alpha_i$ is an $r_i$-th primitive root of unity, for every $i = 1, \ldots, n$. Now, we fix an arbitrary order in $\overline{D}(\mathcal{C})$ and following it, for each $\mathbf{j} \in \mathbb{Z}_{r_1} \times \cdots \times \mathbb{Z}_{r_n}$, we concatenate the vectors $h_e^{\mathbf{j}}$ to get a new vector $h_{\mathbf{j}} = \left( h_e^{\mathbf{j}} \right)_{e \in \overline{D}(\mathcal{C})}$.

Observe that, the equality

$$\sum_{e \in \overline{D}(\mathcal{C})} \gamma_{n+1}(e) = \sum_{e \in \overline{D}(\mathcal{C})} |Q(e)| = |\mathcal{Z}(\mathcal{C})|$$

implies that the length of $h_{\mathbf{j}}$ considered as a vector with entries in $\mathbb{F}_q$ is $|\mathcal{Z}(\mathcal{C})|$. This means that the $\mathbb{F}_q$-subspace spanned by the $h_{\mathbf{j}}$'s verifies $\langle h_{\mathbf{j}} \mid \mathbf{j} \in \mathbb{Z}_{r_1} \times \cdots \times \mathbb{Z}_{r_n} \rangle \subseteq \mathbb{F}_{q^{|\mathcal{Z}(\mathcal{C})|}}$. In fact, as a consequence of Corollary 10, we will see that equality holds.

**Definition 7.** *Let $\mathcal{C} \leq \mathbb{A}(r_1, \ldots, r_n)$ be an abelian code. A check tensor for $\mathcal{C}$ is a set of the form*

$$\{h_{\mathbf{j}} \mid \mathbf{j} \in \mathbb{Z}_{r_1} \times \cdots \times \mathbb{Z}_{r_n}\},$$

*where $h_{\mathbf{j}} = \left(h_e^{\mathbf{j}}\right)_{e \in \overline{D}(\mathcal{C})} \in \mathbb{F}_{q^{|\mathcal{Z}(\mathcal{C})|}}$ and, in turn, $\overline{D}(\mathcal{C})$ is a set of restricted representatives of the $q$-orbits of $D(\mathcal{C})$.*

Note that by definition, $P(X_1, \ldots, X_n) = \sum \lambda_{j_1, \ldots, j_n} X_1^{j_1} \cdots \cdots X_n^{j_n} \in \mathcal{C}$ if and only if $\sum_{\mathbf{j}} \lambda_{\mathbf{j}} h_{\mathbf{j}} = \mathbf{0}$, where $\mathbf{0}$ denotes the zero vector with length $|\mathcal{Z}(\mathcal{C})|$.

**Lemma 8.** *A set $\mathcal{H} \subseteq \mathbb{Z}_{r_1} \times \cdots \times \mathbb{Z}_{r_n}$ is a set of check positions for $\mathcal{C}$ if and only if the set $\{h_{\mathbf{j}}\}_{\mathbf{j} \in \mathcal{H}}$ is a basis of the subspace generated by the check tensor.*

*Proof:* First note that $\mathcal{H}$ is a set of check positions if and only if any codeword is uniquely determined by its entries in the information set $\mathcal{H}^c$, the complementary set of $\mathcal{H}$ in $\mathbb{Z}_{r_1} \times \cdots \times \mathbb{Z}_{r_n}$.

Assume that $\mathcal{H}$ is a set of check positions. Suppose that there exist coefficients $\{\lambda_{\mathbf{j}}\}_{\mathbf{j} \in \mathcal{H}} \subseteq \mathbb{F}_q$ satisfying $\sum_{\mathbf{j} \in \mathcal{H}} \lambda_{\mathbf{j}} h_{\mathbf{j}} = 0$. Define $c = (c_{\mathbf{j}})$ such that $c_{\mathbf{j}} = 0$ if $\mathbf{j} \notin \mathcal{H}$ and $c_{\mathbf{j}} = \lambda_{\mathbf{j}}$ if $\mathbf{j} \in \mathcal{H}$. If one of the coefficients of the linear combination is not zero, one has that $c$ and the zero codeword are different elements of $\mathcal{C}$ and they agree in the positions corresponding to $\mathcal{H}^c$, which is impossible. This proves linear independence. Now, pick $\mathbf{j}_0 \notin \mathcal{H}$. Then, there exists $\{\lambda_{\mathbf{j}}\}_{\mathbf{j} \in \mathcal{H}} \subseteq \mathbb{F}_q$ such that the vector $c = (c_{\mathbf{j}})$, defined by

$$c_{\mathbf{j}} = \begin{cases} 0 & \text{if} \quad \mathbf{j} \in \mathcal{H}^c \setminus \{\mathbf{j}_0\}, \\ 1 & \text{if} \quad \mathbf{j} = \mathbf{j}_0, \\ \lambda_{\mathbf{j}} & \text{if} \quad \mathbf{j} \in \mathcal{H} \end{cases}$$

is a codeword of $\mathcal{C}$. So, $h_{\mathbf{j}_0} + \sum_{\mathbf{j} \in \mathcal{H}} \lambda_{\mathbf{j}} h_{\mathbf{j}} = 0$.

Conversely, suppose that $\{h_{\mathbf{j}}\}_{\mathbf{j} \in \mathcal{H}}$ is linearly independent and for every $\mathbf{i} \in \mathbb{Z}_{r_1} \times \cdots \times \mathbb{Z}_{r_n}$, $h_{\mathbf{i}}$ can be written as an $\mathbb{F}_q$-linear combination of the elements of $\{h_{\mathbf{j}}\}_{\mathbf{j} \in \mathcal{H}}$.

Take any codeword $c = (c_{\mathbf{j}})_{\mathbf{j} \in \mathbb{Z}_{r_1} \times \cdots \times \mathbb{Z}_{r_n}}$ of $\mathcal{C}$. By hypothesis, the vector $\sum_{\mathbf{j} \in \mathcal{H}^c} c_{\mathbf{j}} h_{\mathbf{j}}$ has a unique expression $\sum_{\mathbf{j} \in \mathcal{H}^c} c_{\mathbf{j}} h_{\mathbf{j}} = \sum_{\mathbf{j} \in \mathcal{H}} \lambda_{\mathbf{j}} h_{\mathbf{j}}$, for some $\lambda_{\mathbf{j}} \in \mathbb{F}_q$. As $c \in \mathcal{C}$ then $\lambda_{\mathbf{j}} = -c_{\mathbf{j}}$, for all $\mathbf{j} \in \mathcal{H}$, which means that the codeword is totally determined by those $\{c_{\mathbf{j}}\}_{\mathbf{j} \in \mathcal{H}^c}$. ∎

Now we state our main theorem.

**Theorem 9.** *Let $\mathcal{C} \leq \mathbb{A}(r_1, \ldots, r_n)$ be an abelian code with defining set $D(\mathcal{C})$. Then $\Gamma(\mathcal{C})$ is a set of check positions for $\mathcal{C}$.*

*Proof:* By Lemma 8 it is enough to prove that the set $\{h_{\mathbf{j}}\}_{\mathbf{j} \in \Gamma(\mathcal{C})}$ is a basis of the $\mathbb{F}$-vector space generated by the check tensor; furthermore, we are going to show that it is a basis of the whole space $\mathbb{F}_{q^{|\mathcal{Z}(\mathcal{C})|}}$.

First, we will prove that $|\Gamma(\mathcal{C})| = |\mathcal{Z}(\mathcal{C})|$. For any fixed $(u_n, \ldots, u_2)$, with $1 \leq u_n \leq s_n$ and $1 \leq u_l \leq s(u_n, \ldots, u_{l+1})$, with $2 \leq l < n$, let us define for $i =$

$2, \ldots, n-1$

$$\sigma_i(u_n, \ldots, u_{i+1}) = \sum_{u_i=1}^{s(u_n, \ldots, u_{i+1})} (f[u_n, \ldots, u_{i+1}, u_i] -$$

$$f[u_n, \ldots, u_{i+1}, u_i + 1]) \cdots \sum_{u_2=1}^{s(u_n, \ldots, u_3)} (f[u_n, \ldots, u_3, u_2] -$$

$$f[u_n, \ldots, u_3, u_2 + 1]) \cdot g[u_n, \ldots, u_2].$$

Then, it is clear that

$$|\Gamma(\mathcal{C})| = \sum_{u_n=1}^{s_n} (f[u_n] - f[u_n + 1]) \sigma_{n-1}(u_n)$$

and

$$\sigma_{i+1}(u_n, \ldots, u_{i+2}) = \sum_{u_{i+1}=1}^{s(u_n, \ldots, u_{i+2})} (f[u_n, \ldots, u_{i+2}, u_{i+1}] - \quad (4)$$

$$f[u_n, \ldots, u_{i+2}, u_{i+1} + 1]) \cdot \sigma_i(u_n, \ldots, u_{i+1}).$$

We claim that for $i = 2, \ldots, n-1$ one has

$$\sigma_i(u_n, \ldots, u_{i+1}) = \sum_{e \in \overline{D}_{i-1}(\mathcal{C})} \mu_{u_n, \ldots, u_{i+1}}(e) \cdot \gamma_i(e). \quad (5)$$

We proceed by induction. Taking $i = 2$ as induction base,

$$\sigma_2(u_n, \ldots, u_3) = \sum_{u_2=1}^{s(u_n, \ldots, u_3)} (f[u_n, \ldots, u_3, u_2] -$$

$$f[u_n, \ldots, u_3, u_2 + 1]) \cdot g[u_n, \ldots, u_2].$$

For every $e \in \overline{D}_1(\mathcal{C})$, set $\mu(e) = \mu_{u_n, \ldots, u_3}(e)$. Having in mind that

$$g[u_n, \ldots, u_2] = \sum_{\substack{e \in \overline{D}_1(\mathcal{C}) \\ \mu(e) \geq f[u_n, \ldots, u_2]}} m(e),$$

we factor out $m(e)$ obtaining

$$\sigma_2(u_n, \ldots, u_3) = \sum_{e \in \overline{D}_1(\mathcal{C})} m(e) \cdot$$

$$\sum_{f[u_n, \ldots, u_3, u_2] \leq \mu(e)} (f[u_n, \ldots, u_3, u_2] - f[u_n, \ldots, u_3, u_2 + 1])$$

$$= \sum_{e \in \overline{D}_1(\mathcal{C})} m(e) \cdot \mu(e) = \sum_{e \in \overline{D}_1(\mathcal{C})} \gamma_2(e) \cdot \mu_{u_n, \ldots, u_3}(e).$$

So the induction base is stablished. Now, we assume that the claim has been proved for $i < n-1$ and we are going to deal with the case $i + 1$. By (4) and by assumption

$$\sigma_{i+1}(u_n, \ldots, u_{i+2}) = \sum_{u_{i+1}=1}^{s(u_n, \ldots, u_{i+2})} (f[u_n, \ldots, u_{i+1}] -$$

$$f[u_n, \ldots, u_{i+1} + 1]) \cdot \sum_{e \in \overline{D}_{i-1}(\mathcal{C})} \mu_{u_n, \ldots, u_{i+1}}(e) \cdot \gamma_i(e)$$

$$= \sum_{u_{i+1}=1}^{s(u_n, \ldots, u_{i+2})} (f[u_n, \ldots, u_{i+1}] - f[u_n, \ldots, u_{i+1} + 1]) \cdot$$

$$\sum_{e \in \overline{D}_{i-1}(\mathcal{C})} \gamma_i(e) \cdot \left( \sum_{\mu_{u_n, \ldots, u_{i+2}}(e, a) \geq f[u_n, \ldots, u_{i+1}]} m(e, a) \right).$$

Note that given $e = (e_1, \ldots, e_i) \in \overline{D}_i(\mathcal{C})$, with $\mu_{u_n, \ldots, u_{i+2}}(e) = f[u_n, \ldots, u_{i+2}, l]$ for some $1 \leq l \leq n$, one has that $\pi_{i-1}(e) \in \overline{D}_{i-1}(\mathcal{C})$. Moreover, $e_i$ satisfies that $\mu_{u_n,\ldots,u_{i+2}}(e_1,\ldots,e_{i-1},e_i) \geq f[u_n,\ldots,u_{i+2},u_{i+1}]$ for all $u_{i+1} \leq l$. Conversely, for $e \in \overline{D}_{i-1}(\mathcal{C})$ and $a \in R(e)$, there must exist $u_{i+1}$ such that $\mu_{u_n,\ldots,u_{i+2}}(e,a) \geq f[u_n,\ldots,u_{i+2},u_{i+1}]$. Then, setting $\mu(e) = \mu_{u_n,\ldots,u_{i+2}}(e)$, we have

$$\sigma_{i+1}(u_n,\ldots,u_{i+2}) = \sum_{e \in \overline{D}_i(\mathcal{C})} \gamma_{i+1}(e) \cdot$$
$$\sum_{f[u_n,\ldots,u_{i+1}] \leq \mu(e)} (f[u_n,\ldots,u_{i+1}] - f[u_n,\ldots,u_{i+1}+1])$$
$$= \sum_{e \in \overline{D}_i(\mathcal{C})} \gamma_{i+1}(e) \cdot \mu(e).$$

So (5) is proved. Now, using (5) with $i = n-1$, and the definitions of $\mu_{u_n}(e)$ and $M(e)$, we deduce

$$\begin{aligned}|\Gamma(\mathcal{C})| &= \sum_{u_n=1}^{s_n} (f[u_n] - f[u_n+1]) \cdot \\ &\quad \sum_{e \in \overline{D}_{n-2}(\mathcal{C})} \gamma_{n-1}(e) \cdot \mu_{u_n}(e) \\ &= \sum_{u_n=1}^{s_n} (f[u_n] - f[u_n+1]) \cdot \sum_{e \in \overline{D}_{n-2}(\mathcal{C})} \gamma_{n-1}(e) \\ &\quad \cdot \sum_{M(e,a) \geq f[u_n]} m(e,a) \\ &= \sum_{e \in \overline{D}_{n-1}(\mathcal{C})} \gamma_n(e) \cdot \sum_{f[u_n] \leq M(e)} (f[u_n] - f[u_n+1]) \\ &= \sum_{e \in \overline{D}_{n-1}(\mathcal{C})} \gamma_n(e) \cdot M(e) = \sum_{e \in \overline{D}(\mathcal{C})} \gamma_{n+1}(e).\end{aligned}$$

This equality, together with the definition of $\gamma_{n+1}(e)$ yields

$$|\Gamma(\mathcal{C})| = \sum_{e \in \overline{D}(\mathcal{C})} \gamma_{n+1}(e) = \sum_{e \in \overline{D}(\mathcal{C})} |Q(e_1,\ldots,e_n)| = |\mathcal{Z}(\mathcal{C})|,$$

and we are done.

Now, we have to see that $\{h_{\mathbf{j}}\}_{\mathbf{j} \in \Gamma(\mathcal{C})}$ is linearly independent. So, consider a family $\{\lambda_{\mathbf{j}}\}_{\mathbf{j} \in \Gamma(\mathcal{C})}$ of coefficients in $\mathbb{F}$, in which there is at least one $\lambda_{\mathbf{j}} \neq 0$, for some $\mathbf{j} \in \Gamma(\mathcal{C})$. We shall show that $\sum_{\mathbf{j} \in \Gamma(\mathcal{C})} \lambda_{\mathbf{j}} h_{\mathbf{j}} \neq 0$, and this will finish the proof.

Let us define the set

$$\begin{aligned}\Upsilon &= \{(u_n,\ldots,u_2) \mid 1 \leq u_n \leq s_n, \text{ and} \\ &\quad 1 \leq u_l \leq s(u_n,\ldots,u_{l+1}), \text{ for } 2 \leq l \leq n-1\}\end{aligned}$$

ordered in the usual lexicographical order: $(u_n,\ldots,u_2) \leq (u'_n,\ldots,u'_2)$ if $u_n \leq u'_n$ or if $u_j = u'_j$ for $j = n,\ldots,i+2$ and $u_i \leq u'_i$. Recall that an element $\mathbf{j} = (j_1,\ldots,j_n) \in \mathbb{Z}_{r_1} \times \cdots \times \mathbb{Z}_{r_n}$ belongs to $\Gamma(\mathcal{C})$ if and only if there exists a unique $(u_n,\ldots,u_2) \in \Upsilon$, which we call $\bar{u}_{\mathbf{j}}$, such that $f[u_n,\ldots,u_l+1] \leq j_l < f[u_n,\ldots,u_l]$ for $l = 2,\ldots,n$, and $1 \leq j_1 < g[u_n,\ldots,u_2]$.

We have to prove that there is an element $(e_1,\ldots,e_n) \in \overline{D}(\mathcal{C})$ such that

$$\sum_{(j_1,\ldots,j_n) \in \Gamma(\mathcal{C})} \lambda_{\mathbf{j}} \cdot \alpha_1^{e_1 j_1} \cdots \alpha_n^{e_n j_n} \neq 0.$$

To do this, we begin by fixing $\bar{v}_{\mathbf{i}} = \min\{\bar{u}_{\mathbf{j}} \in \Upsilon \mid \lambda_{\mathbf{j}} \neq 0\}$; where $\mathbf{i} = (i_1,\ldots,i_n) \in \Gamma(\mathcal{C})$. Set $\bar{v}_{\mathbf{i}} = (v_n,\ldots,v_2)$. We shall construct recursively an element $(e_1,\ldots,e_n) \in \overline{D}(\mathcal{C})$ such that for each $l = 1,\ldots,n$,

$$\sum_{\mathbf{j}=(j_1,\ldots,j_l,i_{l+1},\ldots,i_n) \in \Gamma(\mathcal{C})} \lambda_{\mathbf{j}} \cdot \alpha_1^{e_1 j_1} \cdots \alpha_l^{e_l j_l} \neq 0 \quad (6)$$

and while $l < n$,

$$\mu_{v_n,\ldots,v_{l+2}}(e_1,\ldots,e_l) \geq f[v_n,\ldots,v_{l+1}] \text{ if } l \leq n-2, \quad (7)$$
$$M(e_1,\ldots,e_{n-1}) \geq f[v_n] \text{ if } l = n-1. \quad (8)$$

We begin with $l = 1$. By the definition of $\bar{v}_{\mathbf{i}}$ and since $\mathbf{i} \in \Gamma(\mathcal{C})$ we have that $1 \leq i_1 < g[v_n,\ldots,v_2]$. Consider the polynomial

$$P(X_1) = \sum_{\mathbf{j}=(j_1,i_2,\ldots,i_n) \in \Gamma(\mathcal{C})} \lambda_{\mathbf{j}} X_1^{j_1}.$$

By definition of $\Gamma(\mathcal{C})$, we have $\bar{u}_{\mathbf{j}} = \bar{v}_{\mathbf{i}}$, for $\mathbf{j} = (j_1,i_2,\ldots,i_n) \in \Gamma(\mathcal{C})$ and then $j_1$ must verify $1 \leq j_1 < g[v_n,\ldots,v_2]$. Conversely, for any $1 \leq j_1 < g[v_n,\ldots,v_2]$ one has $(j_1,i_2,\ldots,i_n) \in \Gamma(\mathcal{C})$. So $\delta(P(X_1)) = |\{j_1 \mid 1 \leq j_1 < g[v_n,\ldots,v_2]\}| < g[v_n,\ldots,v_2]$, where $\delta$ denote the polynomial degree.

Now, since

$$\begin{aligned}g[v_n,\ldots,v_2] &= \Big| \bigcup \big\{C_{(q,r_1)}(a) \mid a \in \overline{D}_1(\mathcal{C}) \text{ and} \\ &\quad \mu_{v_n,\ldots,v_3}(a) \geq f[v_n,\ldots,v_2]\big\}\Big|\end{aligned}$$

and $P(X_1) \in \mathbb{F}_q[X_1]$ then there exists $e_1 \in \overline{D}_1(\mathcal{C})$ satisfying (7) for $l = 1$, and also $P(\alpha_1^{e_1}) \neq 0$. This proves the induction step $l = 1$.

Suppose we have constructed $e = (e_1,\ldots,e_l) \in \overline{D}_l(\mathcal{C})$ satisfying (6) and (7). We want to get the step $l+1$.

Consider the polynomial

$$P_e(X_{l+1}) = \sum_{\mathbf{j}=(j_1,\ldots,j_{l+1},i_{l+2},\ldots,i_n) \in \Gamma(\mathcal{C})} \lambda_{\mathbf{j}} \cdot \alpha_1^{e_1 j_1} \cdots \alpha_l^{e_l j_l} \cdot X_{l+1}^{j_{l+1}}.$$

Then, for any $\mathbf{j} = (j_1,\ldots,j_{l+1},i_{l+2},\ldots,i_n) \in \Gamma(\mathcal{C})$, we have that $\bar{u}_{\mathbf{j}} = (u_n,\ldots,u_2)$ verifies $u_n = v_n,\ldots,u_{l+2} = v_{l+2}$. By definition of $\bar{v}_{\mathbf{i}}$ we have that $\lambda_{\mathbf{j}} \neq 0$ implies $\bar{u}_{\mathbf{j}} \geq \bar{v}_{\mathbf{i}}$; which, in turn, implies $f[v_n,\ldots,v_{l+1}] \geq f[v_n,\ldots,u_{l+1}]$. Thus, $f[v_n,\ldots,u_{l+1}+1] \leq j_{l+1} < f[v_n,\ldots,u_{l+1}] \leq f[v_n,\ldots,v_{l+1}]$. Then $\delta(P_e(X_{l+1})) < f[v_n,\ldots,v_{l+1}]$ and, by induction hypothesis, we have $f[v_n,\ldots,v_{l+1}] \leq \mu_{v_n,\ldots,v_{l+2}}(e)$.

Analogously to the case $l = 1$ we have that, since

$$\begin{aligned}\mu_{v_n,\ldots,v_{l+2}}(e) &= \Big| \bigcup \big\{C_{(q^{\gamma_{l+1}(e)},r_{l+1})}(a) \mid (e,a) \in \\ &\quad \overline{D}_{l+1}(\mathcal{C}) \text{ and } \mu_{v_n,\ldots,v_{l+3}}(e,a) \geq f[v_n,\ldots,v_{l+2}]\big\}\Big|\end{aligned}$$

and $P_e(X_{l+1}) \in \mathbb{F}_{q^{\gamma_{l+1}(e)}}[X_{l+1}]$, then there exists $e_{l+1} \in \mathbb{Z}_{r_{l+1}}$ satisfying (6) and (7). Now we may get the step $l = n-1$ in a complete analogous way obtaining (8), instead of (7).

Finally, suppose we have an element $e = (e_1,\ldots,e_{n-1}) \in \overline{D}_{n-1}(\mathcal{C})$, satisfying (6), (7) and (8). Consider the polynomial

$$P_e(X_n) = \sum_{\mathbf{j}=(j_1,\ldots,j_n) \in \Gamma(\mathcal{C})} \lambda_{\mathbf{j}} \cdot \alpha_1^{e_1 j_1} \cdots \alpha_l^{e_{n-1} j_{n-1}} \cdot X_n^{j_n}.$$

As in the previous steps, if $\lambda_{\mathbf{j}} \neq 0$ then $\bar{u}_{\mathbf{j}} \geq \bar{u}_{\mathbf{i}}$; hence $u_n \geq v_n$. This means that $f[u_n] \leq f[v_n]$, and since $f[u_n+1] \leq j_n < f[u_n] \leq f[v_n]$, we have $\delta(P_e(X_n)) < f[v_n] \leq M(e) = \sum_{a \in R(e)} m(e,a)$. Thus, there is an $e_n \in \mathbb{Z}_{r_n}$ such that $(e_1,\ldots,e_n) \in \overline{D}(\mathcal{C})$ and $P_e(\alpha_n^{e_n}) \neq 0$; that is

$$\sum_{\mathbf{j}=(j_1,\ldots,j_n) \in \Gamma(\mathcal{C})} \lambda_{\mathbf{j}} \cdot \alpha_1^{e_1 j_1} \cdots \alpha_n^{e_n j_n} \neq 0$$

as desired. ∎

The following corollary is well-known (see for instance [6]).

**Corollary 10.** *Let $\mathcal{C}$ be an abelian code with defining set $D(\mathcal{C})$. Suppose $\mathcal{C}$ has dimension $k$ and length $l$. Then $k = l - |\mathcal{Z}(\mathcal{C})|$.*

## VI. APPLICATIONS TO PERMUTATION DECODING

In this section we present some ideas on how one may design abelian codes with good information sets in order to make a permutation decoding attempt. By using those information sets obtained in Section III, we present two examples. The first example has two parts. In the first one, we give a list of 2-error correcting cyclic codes of length 45 that achieve the upper bound on the ranks that appears in [13]. Next, we construct two 2-error correcting abelian *non cyclic* codes of length 45 with better rates than those given for cyclic ones. The other example is a list of 3-error correcting cyclic codes of length 65 that improve the upper bound on the ranks that appears in [14].

Before we present our examples let us give a brief introduction to the permutation decoding algorithm. Permutation decoding was introduced by F. J. MacWilliams in [9] and it is fully described in [5] and [10]. Fixed an information set for a given linear code $\mathcal{C}$, this technique uses a special set of permutation automorphisms of the code called PD-set.

We denote the permutation group on $\mathbb{Z}_{r_1} \times \cdots \times \mathbb{Z}_{r_n}$ by $S_{r_1 \times \cdots \times r_n}$ and we consider its extension to automorphisms of $\mathbb{A}(r_1, \ldots, r_n)$ via $\tau\left(\sum_{\mathbf{j}} a_{\mathbf{j}} X^{\mathbf{j}}\right) = \sum_{\mathbf{j}} a_{\tau^{-1}(\mathbf{j})} X^{\mathbf{j}}$. Under this point of view the permutation automorphism group of an abelian code $\mathcal{C} \leq \mathbb{A}(r_1, \ldots, r_n)$ is $\mathrm{PAut}(\mathcal{C}) = \{\tau \in S_{r_1 \times \cdots \times r_n} \mid \tau(\mathcal{C}) = \mathcal{C}\}$.

**Definition 11.** *Let $\mathcal{C}$ be a $t$-error-correcting $[l, k]$ code. Let $\mathcal{I}$ be an information set for $\mathcal{C}$. For $s \leq t$ a $s$-PD-set for $\mathcal{C}$ and $\mathcal{I}$ is a subset $P \subseteq \mathrm{PAut}(\mathcal{C})$ such that every set of $s$ coordinate positions is moved out of $\mathcal{I}$ by at least one element of $P$ (see [8], [9]).*

The idea of permutation decoding is to apply the elements of the PD-set to the received vector until the errors are moved out of the fixed information set. The following theorem shows how to check that the information symbols of a codeword with weight less or equal than $t$ are correct. We denote the weight of a vector $v \in \mathbb{F}^l$ with $wt(v)$.

**Theorem 12** ([5], Theorem 8.1). *Let $\mathcal{C}$ be a $t$-error-correcting $[l, k]$ code with parity check matrix $H$ in standard form. Let $r = c + e$ be a vector, where $c \in \mathcal{C}$ and $wt(e) \leq t$. Then the information symbols in $r$ are correct if and only if $wt\left(Hr^T\right) \leq t$.*

Then, once we have found a PD-set $P \subseteq \mathrm{PAut}(\mathcal{C})$ for the given code $\mathcal{C}$ and information set $\mathcal{I}$, the algorithm of permutation decoding is as follows: take a check-matrix $H$ for $\mathcal{C}$ in standard form. Suppose that we receive a vector $r = c + e$ with $wt(e) \leq t$. Then we calculate the syndromes $H\left(\tau(r)\right)^T$, with $\tau \in P$, until we obtain a vector $H\left(\tau_0(r)\right)^T$ with weight less than or equal to $t$. By the previous theorem, the information symbols of the permuted vector $\tau(r)$ are correct, so by using the parity check equations we obtain the redundancy symbols and then we can construct a codeword $c'$. Finally, we decode to $\tau^{-1}(c') = c$.

In general to find $t$-PD-sets for a given $t$-error correcting code is not at all an easy problem. It depends on the chosen information set and they need not even exist. Moreover, it is clear that the algorithm is more efficient the smaller the size of the PD-set.

Let $T_s$ be the transformation from $\mathbb{A}(r_1, \ldots, r_n)$ into itself, given by $T_s(P(X_1, \ldots, X_n)) = X_s \cdot P(X_1, \ldots, X_n)$, for $s = 1, \ldots, n$. Then it is clear that $T_s$ can be seen as a permutation in $S_{r_1 \times \cdots \times r_n}$, via $T_s(i_1, \ldots, i_n) = (i_1, \ldots, i_s + 1, \ldots, i_n)$ and as such, $\langle \{T_s\}_{s=1}^n \rangle$ may be viewed as a subgroup of $\mathrm{PAut}(\mathcal{C})$ for every abelian code $\mathcal{C} \leq \mathbb{A}(r_1, \ldots, r_n)$. On the other hand, we consider the Frobenius automorphism of $\mathbb{F} = \mathbb{F}_q$ acting on $\mathbb{A}(r_1, \ldots, r_n)$ via $\sigma\left(\sum_{\mathbf{j}} a_{\mathbf{j}} X^{\mathbf{j}}\right) = \sum_{\mathbf{j}} a_{\mathbf{j}} X^{q \cdot \mathbf{j}}$ (recall that $\gcd(r_i, q) = 1$, for all $i = 1, \ldots, n$). As well as those $T_s$'s, the Frobenius automorphism belongs to $\mathrm{PAut}(\mathcal{C})$, for every abelian code $\mathcal{C} \leq \mathbb{A}(r_1, \ldots, r_n)$. We shall look for PD-sets contained in the subgroup $\Lambda = \langle \{T_s\}_{s=1}^n \cup \{\sigma\}\rangle$.

It is well-known that for every cyclic code of dimension $k$, any $k$ consecutive positions form an information set (see [10]). In [13], [14] Shiva, Fung and Tan gave upper bounds on the rank of cyclic codes of certain lengths, that are permutation decodable with respect to the information set mentioned above and with PD-sets contained in $\Lambda$. For our purposes, we shall interpret cyclic codes as abelian codes (in several variables), as it was done in Remarks 2(c). Then we design abelian codes such that $\Lambda$ contains a PD-set with respect to the information set defined in (2) and their rates achieve and even improve those given by Shiva, Fung and Tan.

We only consider abelian codes in two variables; however the procedure may be extended to more variables. So, in the sequel, $\mathcal{C} \leq \mathbb{A}(r_1, r_2)$ will be an abelian code with set of check positions $\Gamma(\mathcal{C})$ given by (3), and $Q(e_1, e_2)$ will denote the $q$-orbit of $(e_1, e_2) \in \mathbb{Z}_{r_1} \times \mathbb{Z}_{r_2}$ as in (1).

We begin by dealing with 2-PD-sets. In this case, if $\Gamma(\mathcal{C})$ intersects all $q$-orbits in $\mathbb{Z}_{r_1} \times \mathbb{Z}_{r_2}$, we may move the support of an 2-error vector into $\Gamma(\mathcal{C})$ by using the subgroup $\Lambda$. Indeed, take $e \in \mathbb{A}(r_1, r_2)$ an error vector with $supp(e) = \{p_1, p_2\} \subseteq \mathbb{Z}_{r_1} \times \mathbb{Z}_{r_2}$ and set $p_j = (p_j^1, p_j^2)$ for $j = 1, 2$. Then use $\langle T_1, T_2 \rangle$ to transform $p_1 \mapsto (0, 0)$ and $p_2 \mapsto p$, for some $p \in \mathbb{Z}_{r_1} \times \mathbb{Z}_{r_2}$. Finally, as there exists an integer $i$ such that $q^i \cdot p \in \Gamma(\mathcal{C})$, then $\sigma^i(0, 0) = (0, 0)$ and $\sigma^i(p)$ belong to $\Gamma(\mathcal{C})$. Summarizing

**Lemma 13.** *Let $\mathcal{C} \leq \mathbb{A}(r_1, r_2)$ be an abelian $t$-error correcting code with set of check positions $\Gamma(\mathcal{C})$. If $t \geq 2$ and $\Gamma(\mathcal{C})$ intersects all $q$-orbits in $\mathbb{Z}_{r_1} \times \mathbb{Z}_{r_2}$, then $\Lambda$ is a 2-PD-set with respect to $\Gamma(\mathcal{C})^c$.*

**Example 14.** We shall design binary 2-error correcting codes with the highest dimension in length 45, such that $\Lambda$ contains a 2-PD-set with respect to our information set.

Firts of all, let us note that it is possible to check, by using the GAP program, that any abelian code (cyclic or not), $\mathcal{C}$, of length 45 with $\dim(\mathcal{C}) \geq 33$ has minimum distance $d(\mathcal{C}) \leq 4$.

We begin with cyclic codes. Consider all 2-orbits in $\mathbb{Z}_5 \times \mathbb{Z}_9$.

They are

$$Q(0,0),\ Q(0,1),\ Q(0,3),\ Q(1,0),\ Q(1,1),\ Q(1,2),$$
$$Q(1,3),\ Q(1,6).$$

We consider the 29-dimensional codes $\mathcal{C}_1, \ldots, \mathcal{C}_6$ with defining sets

$$
\begin{aligned}
D\left(\mathcal{C}_1\right) &= Q(1,2) \cup Q(1,6) \\
D\left(\mathcal{C}_2\right) &= Q(1,1) \cup Q(1,6) \\
D\left(\mathcal{C}_3\right) &= Q(1,2) \cup Q(1,3) \\
D\left(\mathcal{C}_4\right) &= Q(1,1) \cup Q(1,3) \\
D\left(\mathcal{C}_5\right) &= Q(1,0) \cup Q(1,2) \\
D\left(\mathcal{C}_6\right) &= Q(1,0) \cup Q(1,1).
\end{aligned}
$$

A direct computation shows that $d\left(\mathcal{C}_i\right) = 5$ for $i = 1, \ldots, 6$. Since

$$
\Gamma(\mathcal{C}_i) \supset \{(0,0),(0,1),(0,3),(1,0),(1,1),(1,2),
$$
$$
(1,3),(2,3)\}
$$

for $i = 1, \ldots, 6$ then by Lemma 13 we have that $\Lambda$ is a 2-PD-set for $\mathcal{C}_i$ with respect to $\Gamma(\mathcal{C}_i)^c$ for $i = 1, \ldots, 6$.

An easy inspection of the 2-orbits shows that $\mathcal{C}_1, \ldots, \mathcal{C}_6$ are the highest dimensional 2-permutation decodable codes. So, by using as information set $\Gamma(\mathcal{C})^c$, we have achieved the upper bound for length 45 given in [13] with respect to the usual information set.

Now we shall construct two binary 2-error correcting non cyclic codes of length 45. The first one $\mathcal{C}_7$, having dimension $\dim(\mathcal{C}_7) = 31$ will be permutation decodable with PD-set $\langle\{T_1, T_2\}\rangle$ with respect to the information set $\Gamma(\mathcal{C}_7)^c$. The second one, $\mathcal{C}_8$, having dimension $\dim(\mathcal{C}_8) = 32$ will be permutation decodable with PD-set $\Lambda$ with respect to the information set $\Gamma(\mathcal{C}_8)^c$.

In this case, we consider all 2-orbits in $\mathbb{Z}_3 \times \mathbb{Z}_{15}$. They are

$$Q(0,0),\ Q(0,1),\ Q(0,3),\ Q(0,5),\ Q(0,7),$$
$$Q(1,0),\ Q(1,1),\ Q(1,2),\ Q(1,3),\ Q(1,5),$$
$$Q(1,6),\ Q(1,7),\ Q(1,10),\ Q(1,11).$$

One may check that the code $\mathcal{C}_7$ having defining set

$$D(\mathcal{C}_7) = Q(0,3) \cup Q(0,7) \cup Q(1,0) \cup Q(1,11)$$

has dimension $\dim(\mathcal{C}_7) = 31$ and minimum distance $d(\mathcal{C}_7) = 6$, so that, it is a 2-error correcting abelian code. The parameters for $\Gamma(\mathcal{C}_7)$ are $f_{[1]} = 8 > f_{[2]} = 3$, $g_{[1]} = 1 < g_{[2]} = 3$ and they yield

$$
\Gamma(\mathcal{C}_7) = \{(0,0),(0,1),(0,2),(0,3),(0,4),(0,5),
$$
$$
(0,6),(0,7),(1,0),(2,0)\}.
$$

We claim that $\langle T_1,\ T_2\rangle$ is a 2-PD-set with respect to $\Gamma(\mathcal{C}_7)^c$. Take $e \in \mathbb{A}(3, 15)$ an error vector with $supp(e) = \{p_1, p_2\} \subseteq \mathbb{Z}_3 \times \mathbb{Z}_{15}$ and set $p_j = (p_j^1, p_j^2)$ for $j = 1, 2$. Clearly, we may use $\langle T_1, T_2\rangle$ to transform $p_1 \mapsto (x_1, 0)$ and $p_2 \mapsto (x_2, y_2)$, where $x_1, x_2 \in \mathbb{Z}_3$ and $0 \leq y_2 \leq 7$. Then, use $\langle T_1 \rangle$ to transform $(x_2, y_2) \mapsto (0, y_2)$, and we are done.

Note that the same argument may be applied to, for example, the code

$$D(\mathcal{C}) = Q(0,3) \cup Q(0,7) \cup Q(1,0)$$

that has dimension $\dim(\mathcal{C}) = 35$; however, as we have mentioned, its minimum distance cannot be greater than 4.

Now we find an upper bound for dimensions when a PD-set includes the Frobenius automorphism. Consider the code $\mathcal{C}_8$ with defining set

$$D(\mathcal{C}_8) = Q(0,0) \cup Q(1,3) \cup Q(1,7) \cup Q(1,11).$$

One may check that $d(\mathcal{C}_8) = 6$ and $\dim(\mathcal{C}_8) = 32$, so it is a 2-error correcting abelian code. In this case, Lemma 13 does not may be applied; however, we shall see that $\Lambda$ is a PD-set. Take $e \in \mathbb{A}(3, 15)$ an error vector with $supp(e) = \{p_1, p_2\} \subseteq \mathbb{Z}_3 \times \mathbb{Z}_{15}$ and set $p_j = (p_j^1, p_j^2)$ for $j = 1, 2$. By analyzing the distribution of the 2-orbits in the plane $\mathbb{Z}_3 \times \mathbb{Z}_{15}$ we see that, first, one may use $\Lambda$ to transform $p_1 \mapsto (0, 0)$ and $p_2 \mapsto (x, y)$, with $12 \leq y \leq 14$. Then $\langle T_1, T_2\rangle$ finishes the task.

Now we deal with 3-PD-sets.

**Lemma 15.** *Let* $\mathcal{C} \leq \mathbb{A}(r_1, r_2)$ *be an abelian* $t$-*error correcting code with set of check positions* $\Gamma(\mathcal{C})$ *given by (3). If* $t \geq 3$, $f_{[1]} = r_2$, $g_{[s_n]} = r_1$ *and* $\Gamma(\mathcal{C})$ *intersects all* $q$-*orbits in* $\mathbb{Z}_{r_1} \times \mathbb{Z}_{r_2}$, *then* $\Lambda$ *is a 3-PD-set with respect to* $\Gamma(\mathcal{C})^c$.

*Proof:* Take $e \in \mathbb{A}(r_1, r_2)$ an error vector with $supp(e) = \{p_1, p_2, p_3\} \subseteq \mathbb{Z}_{r_1} \times \mathbb{Z}_{r_2}$ and set $p_j = (p_j^1, p_j^2)$ for $j = 1, 2, 3$. Then use $\langle T_1, T_2\rangle$ to transform $p_1 \mapsto (a_1, 0), p_2 \mapsto (0, a_2)$ and $p_3 \mapsto p$, for some $a_1 \in \mathbb{Z}_{r_1}, a_2 \in \mathbb{Z}_{r_2}$ and $p \in \mathbb{Z}_{r_1} \times \mathbb{Z}_{r_2}$. Now, by hypothesis we have that there exists an integer $i$ such that $q^i \cdot p \in \Gamma(\mathcal{C})$, then $\sigma^i(a_1, 0) = (*, 0)$, $\sigma^i(0, a_2) = (0, *)$ and $\sigma^i(p)$ belong to $\Gamma(\mathcal{C})$. ∎

**Example 16.** We shall design binary 3-error-correcting cyclic codes with the highest dimension in length 65 such that $\Lambda$ contains a 3-PD-set with respect to our information set. In this case, the 2-orbits in $\mathbb{Z}_5 \times \mathbb{Z}_{13}$ are

$$Q(0,0),\ Q(0,1),\ Q(1,0),\ Q(1,1),\ Q(1,2),\ Q(1,4),\ Q(1,7).$$

We consider the 40-dimensional codes $\mathcal{C}_1, \ldots, \mathcal{C}_4$ with defining sets

$$
\begin{aligned}
D\left(\mathcal{C}_1\right) &= Q(0,0) \cup Q(0,1) \cup Q(1,1) \\
D\left(\mathcal{C}_2\right) &= Q(0,0) \cup Q(0,1) \cup Q(1,2) \\
D\left(\mathcal{C}_3\right) &= Q(0,0) \cup Q(0,1) \cup Q(1,4) \\
D\left(\mathcal{C}_4\right) &= Q(0,0) \cup Q(0,1) \cup Q(1,7).
\end{aligned}
$$

A direct computation shows that $d\left(\mathcal{C}_i\right) = 8$ for $i = 1, \ldots, 4$. Since

$$\{(0,0),(0,1),(1,0),(1,1),(1,2),(3,2),(2,1)\} \subset \Gamma(\mathcal{C}_i)$$

for $i = 1, \ldots, 4$ then by Lemma 15 we have that $\Lambda$ is a 3-PD-set for $\mathcal{C}_i$ with respect to $\Gamma(\mathcal{C}_i)^c$ for $i = 1, \ldots, 4$.

An easy inspection of the 2-orbits shows that $\mathcal{C}_1, \ldots, \mathcal{C}_4$ are the highest dimensional 3-permutation decodable codes. So, by using as information set $\Gamma(\mathcal{C})^c$, we have improved the upper bound for length 65 (which is dimension 38) given in [14, Table X] with respect to the usual information set.

## Acknowledgement

The authors sincerely thank professor Ángel del Río for his helpful comments and suggestions.